# High sensitivity single-axis vector magnetic field sensing via a sandwich-type PDMS resonator

WEIKANG XU[1], JIAMIN RONG,[2, *] ENBO XING[1], TAO JIA[1], JIANGLONG LI[1], JUN YUE[1], JUN TANG[2, *], AND JUN LIU[1]

[1]Key Laboratory of Electronic Testing Technology, School of Instrument and Electronics, North University of China, Taiyuan, 030051, China

[2] School of Semiconductors and Physics, North University of China, Taiyuan, 030051, China

*Corresponding author: rongjiamin@126.com, tangjun@nuc.edu.cn

**The sandwich structure as the core layer of PDMS resonator is proposed for single-axis magnetic sensor with high sensitivity. The small Young's modulus of flexible material corresponds to larger variation, resulting in a highly sensitive magnetic response. The sandwich structure pre-set magnetic field provides directional sensing feature. The experimental results show that the redshift sensitivity of 1.08 nm/mT and the blueshift sensitivity of 1.12 nm/mT in the unshielded environment, which is attributed to slight variation in PDMS Young's modulus. At 1.4 kHz, a minimum detectable magnetic field of 0.96 nT·Hz$^{-1/2}$ is realized.**

Magnetic sensors have become an essential part in various fields including earthquake prediction, space guidance and navigation, geological exploration, military and medical health [1-4]. Compared with traditional electronic magnetometers, optical magnetic sensors have the advantages of good reliability, wide dynamic range, large spectral bandwidth, strong anti-interference capability and easy integration, such as atomic magnetometers, NV color center optical magnetic sensors and whispering gallery mode (WGM) resonator based optical magnetic sensors [5-11]. In the last decade, optical magnetic sensors based on WGM resonators have a promising application in high-sensitivity magnetic field detection since they have a high-quality factor (Q-factor) and compact mode volume, which can greatly increase the intensity of the optical field and material interaction inside the cavity.

Most magnetic sensors based on WGM resonators are divided into two main types: one type is the frequency shift caused by an external magnetic field in the optical frequency range. Usually, the microbubble is blown inside a hollow capillary tube and filled with magnetic fluid [12-16]. As the external magnetic field increases, the refractive index around the WGM resonator causes a shift in the WGM resonant frequency. The method of resonance wavelength shift affects the detection limit [10].

Another type is the conversion of optical frequency range to mechanical frequency range by photon-phonon coupling. For instance, the WGM resonator is combined with magnetostrictive materials [17-24], and the external magnetic field induces mechanical force to change of transmission spectrum while being able to maintain a high Q-factor. The magnetic-field induced strain of most magnetostrictive materials is very small, usually less than several ppm/mT [25], and bandwidth and sensitivity are limited by the extent to which magnetostrictive materials and WGM resonators are combined, and achieving low noise and high sensitivity simultaneously is a challenge [18].

In this letter, a sandwich structure magnetic sensor based on on-chip PDMS flexible resonator mechanically coupled with a magnet is proposed. The optimal sandwich structure is obtained through finite element simulation of magnetic flux density at different pitches. In unshielded environment, redshift sensitivity of 1.08 nm/mT and a blueshift sensitivity of 1.12 nm/mT. By applying the AC signal in the Hz to kHz range, a minimum detectable magnetic field of 0.96 nT·Hz$^{-1/2}$ is achieved at 1.4 kHz. Theoretical simulations and experiments verify that the magnetic sensor has high sensitivity.

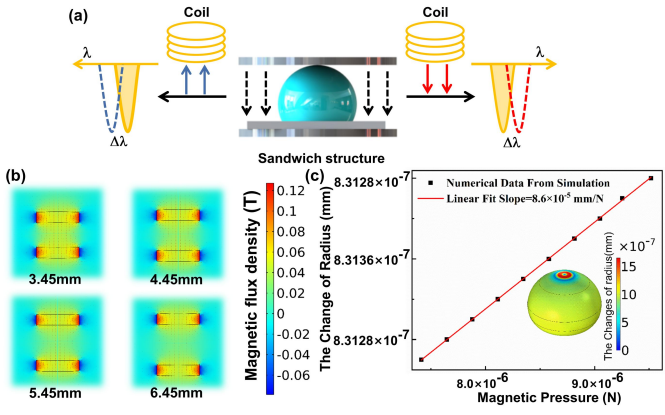

Fig. 1. (a) Principle of magnetic sensing in sandwich structure. (b) Magnetic flux density at different magnet distances. (c) The change of R under different magnetic forces. Inserted is the change of R at a magnetic force of $7.41 \times 10^{-6}$ N.

The sensing principle of the sandwich structure is shown in Fig. 1(a). The WGM resonance condition is $2\pi R n = m\lambda$, where $\lambda$ is the resonance wavelength, R and n are the coupling position radius and effective refractive index of the PDMS flexible resonator, respectively, and m is an integer of the resonance mode order. As shown in Ref. [26, 27], the change in index of refraction is negligible, and the shift in resonance wavelength approximates the change in radius. The magnetic flux density between two neodymium iron boron magnet (NdFeB) magnets (N35) of 8 mm diameter simulated by finite element method is shown in Fig. 1(b). The z-axis force on the 5.45mm sandwich structure is $2.32 \times 10^{-5}$ N/mT. and using these magnetic forces as model inputs, Fig. 1(c) shows that the resonator equatorial radius R shows a linear relationship with a slope of $8.6 \times 10^{-5}$ mm/N, where dB is the variation of the magnetic field, for a magnet spacing of 5.45 mm with different magnetic forces. Thus, the sensitivity of the sandwich structure magnetic sensor can be defined as

$$S = \frac{d\lambda}{dB} = \frac{dR}{dB}\frac{\lambda}{R} = 3.2 \text{nm/mT} \quad (1)$$

The experimental setup for sensing and detection of the PDMS flexible resonator sandwich structure is shown in Fig. 2(a). The incident light from the Toptica laser is passed through an attenuator to keep the incident light at a lower power, which reduces the thermal effect in the PDMS flexible resonator. The transmission spectrum of the sandwich structure is obtained by coupling the 2.1 μm tapered fiber into and out of the sandwich structure. A polarization controller is used to optimize the coupling strength. The light coming out of the tapered fiber is connected to a photodetector and transmitted to an oscilloscope to obtain the transmission spectrum, and a power meter (FMH-87107) is used for data acquisition and mode analysis. By attaching the tapered fiber to the resonator surface, the purpose is to ensure a stable coupling state during the whole experiment and to reduce the effect of ambient fluctuation noise.

Fig. 2(b) is the process of making PDMS flexible resonators. The slides coated with superhydrophobic material are placed at room temperature for 12h. The configuration process of PDMS (Sylgard184) is as follows: the polymer and curing agent are mixed at a weight ratio of 10:1 and stirred well. When the bubbles disappear, the mixed PDMS is aspirated with a syringe and we use the syringe to shape the resonator on the slide coated with superhydrophobic material. The shape of the resonator does not change much, which can ensure the stability of the sandwich structure. Fig. 2(c) and (d) top and side views of the PDMS flexible resonator.

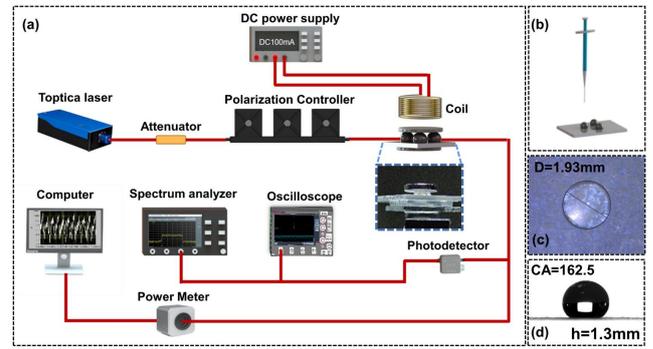

Fig. 2. (a) Schematic diagram of the experimental setup. (b) Schematic diagram of the PDMS flexible resonator fabrication. (c) and (d) top view and side view of the PDMS flexible resonator, respectively. The diameter is 1.93mm, the height is 1.3mm, and the contact angle is 162.5°.

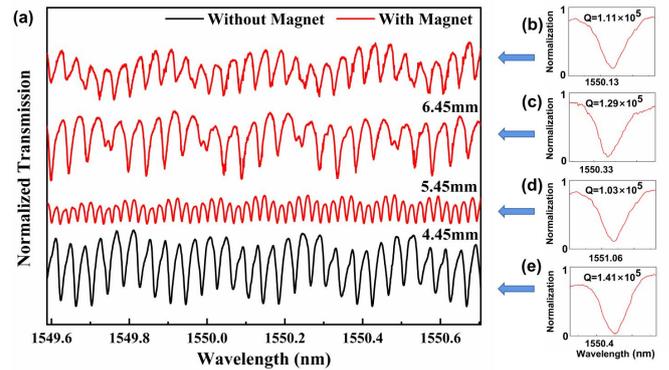

Fig. 3. (a) Transmission spectra of different magnet distances. (b)-(e) Amplified resonance peaks. The Q factors are $1.11 \times 10^5$, $1.29 \times 10^5$, $1.03 \times 10^5$ and $1.41 \times 10^5$, respectively.

Fig. 3(a) shows the transmission spectra of PDMS flexible resonators with sandwich structure at different distances. Fig. 3(b)-(d) show the transmission spectra of the sandwich structure with magnet spacing of 4.45 mm, 5.45 mm, and 6.45 mm, respectively. In Fig. 3(e), the transmission spectrum of the PDMS flexible resonator with $Q = \lambda/d\lambda = 1.41 \times 10^5$ at the resonant wavelength of 1550.426 nm, where dλ is the full width with half-maximum for this optical mode. This demonstrates the ability of the resonator to consistently maintain a Q-factor of $10^5$ under different deformations, providing the basis for subsequent sensing tests.

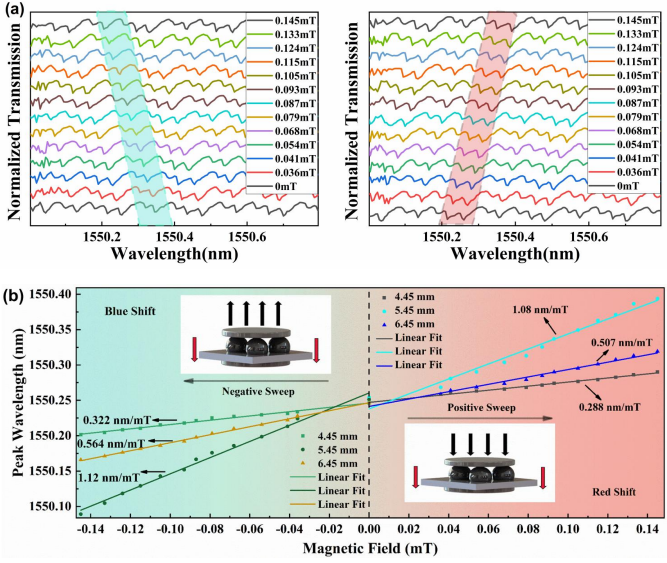

Fig. 4. (a) Transmission spectra under different magnetic fields. (b) Sensitivity at different magnet distances.

The heat insulation coated coil is placed above the sandwich structure. The coil magnetic field signal generation is controlled by a DC power supply, and the supply DC magnetic field range is 0-0.145 mT measured by a Gauss meter. The magnetic field is applied to the sandwich structure with magnet distances of 4.45 mm, 5.45 mm, and 6.45 mm, and the variation of the transmission spectrum is shown in the Fig. 4(a), the shifts of resonance wavelengths are marked with blue area and red area, respectively. When the applied magnetic field direction is the identical to the preset magnetic field direction, the resonance wavelength is redshifted with the increase of magnetic field, and the redshift sensitivity is 0.288 nm/mT, 1.08 nm/mT, 0.507 nm/mT, respectively. Applying the reverse magnetic field, the resonance wavelength is blue-shifted with the increase of magnetic field, and the blueshift sensitivity is 0.322 nm/mT, 1.12 nm/mT, and 0.564 nm/mT, as shown in Fig. 4(b). This result may be caused by the physical properties of the PDMS material. It is found that the blueshift sensitivity is large When the reverse magnetic field direction is applied, the magnetic field between the two magnets is reduced by vector superimposition, the PDMS flexible resonator springs back, and the Young's modulus changes slightly. As the magnetic field slowly increases, the effect on the PDMS Young's modulus is relatively small, and the wavelength shift is linearly proportional to the magnetic field.

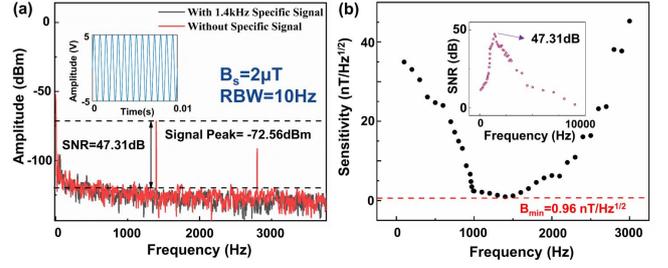

Fig. 5. (a) Power spectral density at a specific frequency of 1.4kHz. (b) Peak sensitivity at different frequencies. Inset: Variation of SNR with frequency.

To test the magnetic field detection capability of this experimental device, we apply a sinusoidal signal from the signal generator to the coil to produce a specific frequency of AC magnetic field signal applied to the sandwich structure, the spectrum analyzer will peak at that specific frequency. Therefore, we use the signal-to-noise ratio (SNR) of the peak signal to calculate the peak sensitivity (magnetic field detection limit), keep the constant amplitude voltage of the signal generator at 5 V, and vary the frequency of the output current to obtain the frequency response of the sandwich structure. By applying a peak magnetic field signal of $B_s$=2 μT at 1.4 kHz and a peak at 1.4 kHz on the spectrum analyzer as shown in Fig. 5 (a), which may be due to the mechanical resonance of the sandwich structure. The SNR was 47.31 dB, and the resolution bandwidth (RBW) is set to 10 Hz. The peak sensitivity at this time at a specific frequency is determined by follow Ref. [19] as

$$B_{min}(\omega_s) = \frac{B_s}{\sqrt{SNR \times RBW}} = 0.96 nT \cdot Hz^{-1/2} \quad \textbf{(2)}$$

Fig. 5(b) shows the peak sensitivity of the system at different frequencies. The SNR varies with the signal field as shown in the inset, with a maximum value at 1.4 kHz, which corresponds to the smallest peak sensitivity at this time. It is possible that this is related to the inherent characteristics of NdFeB magnets, where a smaller magnet may have a larger response to the magnetic field.

In summary, we demonstrate a single-axis magnetic sensor based on a sandwich structure with a PDMS flexible resonator formed on a chip. The magnet and flexible cavity structure provides both the vectoring of the magnetic sensor and excellent magnetic field sensing capability. With the distance of 5.45 mm, the redshift sensitivity and the blueshift sensitivity are respectively 1.08 nm/mT and 1.12 nm/mT. The distinction of sensitivity is related to the slight change in Young's modulus. At 1.4 kHz, a minimum detectable magnetic field of 0.96 nT·Hz$^{-1/2}$ is detected. In addition, the sensor has the advantages of low cost, high sensitivity, good repeatability, and broad potential for unidirectional magnetic field detection.

**Funding.** Joint Funds of the National Natural Science Foundation of China (U21A20141), the Innovative Research Group Project of National Natural Science Foundation of China (51821003), Natural National Science Foundation of China (51922009, 52005457, 62004179), Shanxi Key Laboratory of Advanced Semiconductor Optoelectronic Devices and Integrated Systems (2022SZKF01).

**Disclosures.** The authors declare no conflicts of interest.

**Data availability.** Data underlying the results presented in this paper are not publicly available at this time but may be obtained from the authors upon reasonable request.